\documentstyle[twoside,fleqn,espcrc2,epsf]{article}

\newcommand{\be}{\begin{equation}}
\newcommand{\ee}{\end{equation}}
\newcommand{\bea}{\begin{eqnarray}}
\newcommand{\eea}{\end{eqnarray}}
\newcommand{\pardis}{\langle \mu \rangle}

\title{Dual Superconductivity and Chiral Symmetry in Full QCD}

\author{
    J.M. Carmona\address[ZARA]{Departamento de F\'{\i}sica Te\'orica, 
    Universidad de Zaragoza, 50007 Zaragoza, Spain },
    M. D'Elia\address[GENO]{Dipartimento di Fisica dell'Universit{\`a} di
    Genova and INFN, I-16146, Genova, Italy} \thanks{Speaker at the 
    Conference (Email: delia@ge.infn.it). Partially supported by MURST and 
    by EC, FMRX-CT97-0122}, 
    L. Del Debbio\address[PISA]{Dipartimento di Fisica dell'Universit\`a 
    and INFN, Via Buonarroti 2 Ed. B, I-56127 Pisa, Italy},
    A. Di Giacomo\addressmark[PISA],
    B. Lucini\address[OXFO]{Theoretical Physics, University of Oxford,
    1 Keble Road, OX1 3NP Oxford, UK},
    G. Paffuti\addressmark[PISA]}  

\begin{document}

\begin{abstract}
A disorder parameter detecting dual superconductivity of the vacuum is 
measured across the chiral phase transition in full QCD with two flavours 
of dynamical staggered fermions. The observed behaviour is similar to 
the quenched case.
\end{abstract}

\maketitle

\section{Introduction}
\label{sec:introduction}

Confinement is experimentally a well established property of strong interactions. 
In the quenched approximation a phase transition driven
by temperature from a confined to a deconfined phase has been observed 
by lattice simulations, an order parameter being the v.e.v. of
the Polyakov line, which is zero in the confined (low temperature) phase and nonzero
in the deconfined (high temperature) phase, thus signalling
the spontaneuous breaking of the $Z_3$ centre symmetry.
The situation is different in full QCD, where the presence
of dynamical quarks makes the $Z_3$ centre symmetry explicitely broken,
so that the v.e.v. of the Polyakov line ceases to be an order parameter.
One speaks instead of a phase transition
from a spontaneously broken to a restored chiral symmetry,
the chiral condensate being the order parameter.
Anyway, at the chiral phase transition
a peak in the Polyakov line susceptibility is still observed, 
indicating a sort of deconfining crossover.
However up to now the relation between chiral symmetry
and confinement is still not well understood.

On the other hand, confinement in the quenched theory is now well
understood in terms of an order/disorder phase transition~\cite{'tHooft78,'tHooft81},
with the QCD vacuum behaving like a dual superconductor at low temperatures
and undergoing a phase transition to a normal conducting state at 
the critical temperature $T_c$. A disorder parameter which is
the v.e.v. of a magnetically charged operator has been developed
and used to demonstrate dual superconductivity of the vacuum for
U(1), SU(2) and SU(3) Yang-Mills theory~\cite{ldd,u1,I,II,III}.
Contrary to the v.e.v. of the Polyakov line, the definition of 
the disorder parameter detecting dual superconductivity can be extended
to full QCD. It is then possible to test whether there is any transition from
a dual superconducting to an ordinary vacuum also in presence of dynamical fermions
and how it is related to the chiral phase transition. 
This is the aim of our study and we will present some preliminary
results in this paper. 

\section{The disorder parameter in full QCD}
\label{sec:definition}

The definition of the disorder parameter in pure gauge theories has been 
discussed 
in~\cite{ldd,u1,I,II,III}, 
so we will only recall here the main points.
It is the v.e.v. of a gauge invariant operator, 
$\mu$, which carries magnetic charge. The $U(1)$ magnetic subgroup is selected
by the so called abelian projection. A gauge fixing is performed, usually
by diagonalizing a local operator in the adjoint representation, which leaves
a U(1)$\times$U(1) gauge freedom (for SU(3)), corresponding 
to two possible definitions of the magnetic charge. It has been proved
in~\cite{III} that the dual superconductivity mechanism for confinement
works independently of the abelian projection chosen, as already
suggested in~\cite{'tHooft81}.

On the lattice $\mu$ can be written as~\cite{ldd,u1,I,II,III}
\be
\mu \equiv \exp{ \frac{\beta}{3} \sum_{ i, \vec{x}} {\rm Re}{\rm Tr}
\left( \tilde{U}_{0 i}(\vec{x},t_0) - {U}_{0 i}(\vec{x},t_0) \right)} \ , 
\ee
where the sum in the exponential is extended only to the temporal plaquettes
${U}_{0 i}$ in the time slice $t = t_0$ where the magnetic monopole is created, and
$\tilde{U}_{0 i }({\vec{x},t_0})$ indicates a plaquette modified
by the insertion of a monopole field, defined as an abelian field in
the abelian projected gauge.
The disorder parameter is then defined as
\be
\langle \mu \rangle = 
\frac{\int \left( {\cal D} U \right) \mu e^{-S_G} }
{\int \left( {\cal D}U \right)  e^{-S_G}} \ =
\frac{\int \left( {\cal D} U \right) e^{-S_{M}}}
{\int \left( {\cal D}U \right)  e^{-S_G}} \ ,
\ee
where $S_G$ is the standard Wilson action and $S_M$ is a modified action 
obtained by inserting the monopole field in the temporal plaquettes
on time slice $t_0$. It can be shown~\cite{u1}, by a change of integration 
variables, that the insertion of $\mu$ in the functional integral
corresponds to creating a monopole at $t_0$
which then propagates at all times $t > t_0$. For this reason, periodic
boundary conditions in time direction are not consistent with the definition
of $\pardis$, and $C^{\star}$-periodic boundary conditions have to be used instead: 
\be
U_\mu(\vec{x},t=N_t) = U_\mu^\star (\vec{x},t=0) \ ,
\ee
$N_t$ being the temporal extension of the lattice and $U_\mu^\star$ 
being the complex conjugated of $U_\mu$.

A lattice determination of $\langle \mu \rangle$ is difficult,
since it is the average of a quantity which fluctuates exponentially with the 
square root of the physical volume. It is much more convenient to
study
\be
\rho = \frac{\partial}{\partial \beta} \log \langle \mu \rangle = \langle S_G \rangle_{S_G} -
\langle S_M \rangle_{S_M} \ ,
\label{rhodef}
\ee
where $\langle \cdot \rangle_S$ indicates the v.e.v. obtained using the action
$S$ to weight configurations. From the knowledge of $\rho$ 
the relevant physical information on $\pardis$
can be extracted. The result in pure gauge theory is that $\pardis \neq 0$ 
in the confined, low temperature phase, while it goes to zero at $T_c$, with 
critical indices which can be extracted from the finite size scaling behaviour
of $\rho$.

In full QCD the definitions of $\pardis$ and $\rho$ are modified in a very 
simple way:
\be
\pardis = 
\frac{\int \left( {\cal D} \bar{\psi} {\cal D} \psi {\cal D} U \right) e^{-S_M - S_F} }
{\int \left( {\cal D} \bar{\psi} {\cal D} \psi {\cal D}U \right)  e^{-S_G -S_F}} \ ,
\ee
where $S_F$ is the fermionic action, and
\be
\rho =  \langle S_G \rangle_{S_G + S_F} -
\langle S_M \rangle_{S_M + S_F} \ .
\label{rhoferm}
\ee

Unlike the $Z_3$ centre symmetry, 
the $U(1)$ magnetic symmetry defined after abelian projection is 
still a good symmetry also in presence of dynamical fermions.
Therefore $\pardis$ can be a correct disorder parameter
for the dual superconductivity transition also in full QCD.

\section{Algorithm implementation and simulation details}
\label{sec:details}

We have used two flavours of staggered fermions and the R version of 
the HMC algorithm for our simulations. 
Some technical complications arise in the computation
of the second member on the right hand side of Eq. (\ref{rhoferm}).
In the evaluation of 
$\langle S_M \rangle_{S_M + S_F}$, the use of $C^{\star}$-periodic boundary
conditions in time direction for the gauge fields requires
$C^{\star}$ boundary conditions in
temporal direction also for fermionic variables 
(in addition to the usual antiperiodic ones), in order to ensure gauge 
invariance of the fermionic determinant.
This implies relevant changes in the formulation
and implementation of the HMC algorithm which are explained in detail
in Ref.~\cite{cstar}.

We have chosen the Polyakov line as the local adjoint operator which 
defines the abelian projection. Actually, calling $L(\vec{x},t)$ the Polyakov 
line starting at point $(\vec{x},t)$, the abelian projection is defined
by the operator $L(\vec{x},t) L^\star(\vec{x},t)$, which transforms 
in the adjoint representation when using $C^\star$ boundary 
conditions.

The use of a modified gauge action also implies changes in the 
molecular dynamics equations. One has to maintain 
constant the modified hamiltonian containing $S_M$. 
A change in any temporal link indeed induces
a change in $L(\vec{x},t)$ and hence in the abelian projection 
defining the monopole field. Therefore the dependence of $S_M$ on 
temporal links is non trivial and the equations of motion
for the temporal momenta become more complicated. Details
of the changes necessary in the molecular dynamics equations
will be published in a forthcoming paper~\cite{tocome}.
\begin{figure}[t]
\vspace{-4pt}
\begin{center}
\leavevmode
\epsfxsize=66mm
\epsffile{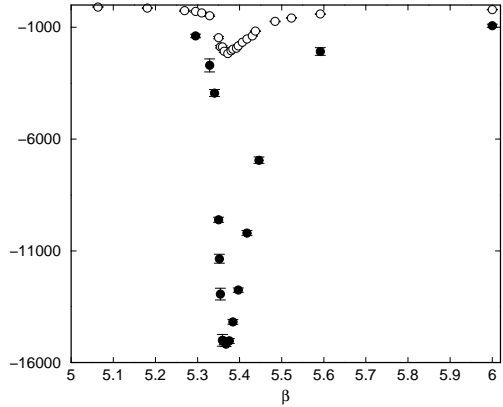}
\end{center}
\vspace{-35pt}
\label{fig1}
\caption{Results for $\rho$ on the $16^3 \times 4$ (open circles) and 
the $32^3 \times 4$ (filled circles) lattice.}
\vspace{-18pt}
\end{figure}
We have used a temporal lattice extent $N_t = 4$
and spatial extents $N_s = 16,32$.
We have chosen to vary the temperature,
$T = 1/( N_t a(\beta, m_q))$, moving in the $(\beta, m_q)$ plane
while keeping a fixed value of $m_\pi/m_\rho$. To do this
and to extract the physical scale we have used fits to the 
$m_\rho$ and $m_\pi$ masses published in~\cite{blum}. In particular
we present here results obtained at $m_\pi/m_\rho \simeq 0.505$.

\section{Results and Outlook}
\label{sec:results}

Results obtained for $\rho$ on the $16^3 \times 4$ and $32^3 \times 4$
lattices are shown in Fig.~1. A clear peak is present in both cases and, as can be
seen in Fig.~2, it is located exactly at the chiral phase
transition, at a critical temperature $T_c \sim 150$ MeV.
Moreover, as is visible in Fig.~1, the peak scales roughly by a factor
of eight when going from $N_s = 16$ to $N_s = 32$, i.e. $\propto N_s^3$.
Previous studies of full QCD at a similar set of 
physical parameters~\cite{jlqcd} have found a crossover rather than a 
phase transition at $T_c$.
\begin{figure}[t]
\vspace{-4pt}
\begin{center}
\leavevmode
\epsfxsize=72mm
\epsffile{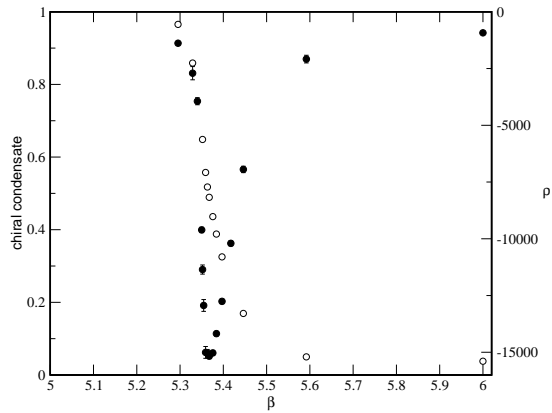}
\end{center}
\vspace{-35pt}
\label{fig2}
\caption{Chiral condensate (open circles) and $\rho$ (filled circles)
on the $32^3 \times 4$ lattice.}
\vspace{-14pt}
\end{figure}
From our present results we can draw the following conclusions:
{\it i)} the disorder parameter shows a behaviour very similar to 
that observed in pure Yang-Mills theories, thus indicating the presence
of a transition from a dual superconducting to an ordinary
vacuum state also for full QCD. This transition clearly happens
in correspondence of the chiral phase transition; {\it ii)} our results
seem to scale roughly with a critical index
$\nu \sim 1/3$. Of course we still need to collect more data near
the critical point and at different values of $N_s$ in order 
to draw more precise conclusions about the critical behaviour of 
$\rho$ at $T_c$.

\end{document}